\documentstyle[epsfig,aps,prb]{revtex}
\begin{document}
\draft
\preprint{1st-draft}
\title{
The "devil's staircase" type phase transition in NaV$_{2}$O$_{5}$ under high pressure
}

\author{K. Ohwada\cite{preadd}, Y. Fujii, N. Takesue, 
M. Isobe$^{1}$, Y. Ueda$^{1}$, H. Nakao$^{2}$, 
Y. Wakabayashi$^{2,3,}$\cite{preadd2},\\ Y. Murakami$^{2,}$\cite{preadd3}, K. Ito$^{4}$, Y. Amemiya$^{5}$, Y. Fujihisa$^{6}$,
K. Aoki$^{6}$, T. Shobu$^{7,}$\cite{preadd4}, Y. Noda$^{7}$ and N. Ikeda$^{8}$
}

\address{
Neutron Scattering Laboratory, Institute for Solid States Physics, The
University of Tokyo,\\
106-1 Shirakata, Tokai, Ibaraki 319-1106, Japan\\
$^{1}$Material Design and Characterization Laboratory, Institute for
Solid States Physics,\\ The University of Tokyo,
5-1-5 Kashiwanoha, Kashiwa, Chiba 277-8581, Japan\\
$^{2}$Photon Factory, Institute of Materials Structure Science, High
Energy Accelerator Research Organization (KEK),\\
1-1 Oho, Tsukuba, Ibaraki 305-0801, Japan\\
$^{3}$Science and Technology, Keio University, 3-14-1, Hiyoshi, Kouhoku-ku, Yokohama, 223-8522   Japan\\
$^{4}$Institute of Material Science, University of Tsukuba,\\
1-1-1 Tennoudai, Tsukuba, Ibaraki 305-8577, Japan\\
$^{5}$Department of Advanced Materials Science, Graduate School of Frontier Science,\\ The University of Tokyo, 7-3-1 Hongo, Bunkyo-ku, Tokyo 113-8656, Japan\\
$^{6}$National Institute of Materials and Chemical Research, Agency of
Industrial Science and Technology,\\
1-1 Higashi, Tsukuba, Ibaraki 305-8565, Japan\\
$^{7}$Institute of Multidisciplinary Research for Advanced Materials, Tohoku University,\\
2-1-1 Katahira, Aoba-ku, Sendai, Miyagi 980-8577, Japan\\
$^{8}$The Japan Synchrotron Radiation Research Institute (JASRI), SPring-8,\\
1-1-1 Kouto, Mikazuki, Hyogo 679-5198, Japan\\
}

\date{\today}

\twocolumn[\hsize\textwidth\columnwidth\hsize\csname @twocolumnfalse\endcsname
\maketitle
\newpage

\begin{abstract}
The "devil's staircase" type phase transition in the quarter-filled spin-ladder compound NaV$_{2}$O$_{5}$ has been discovered at low temperature and high pressure by synchrotron radiation x-ray diffraction. A large number of transitions are found to successively take place among higher-order commensurate phases with $2a {\times} 2b {\times} zc$ type superstructures. The observed temperature and pressure dependence of modulation wave number q$_{c}$, defined by $1/z$, is well reproduced by the Axial Next Nearest Neighbor Ising (ANNNI) model. The q$_{c}$ is suggested to reflect atomic displacements coupled with charge ordering in this system. The experimental fact implies that two competitive inter-layer interactions between the Ising spins, i.e., the nearest neighbor $J_{1}>0$ (ferro) and the next nearest neighbor $J_{2}<0$ (antiferro) along the $c$-axis, are intrinsic in this compound. A microscopic origin of the inter-layer interaction is not yet known; however, the nearest neighbor interaction $J_{1}>0$ between the V$_{2}$O$_{5}$ layers is especially interesting. It is very surprising that the phase transition of such a complicated charge-lattice-spin coupled system NaV$_{2}$O$_{5}$ can be described by the simple ANNNI model.
\end{abstract}
]

\narrowtext
%Introduction
Since discovery of spin-Peierls-like behavior in NaV$_{2}$O$_{5}$ by Isobe and Ueda~\cite{isobe1}, a wide variety of experimental and theoretical studies have been carried out.
The recent structure analysis~\cite{new-xray1,new-xray2}, NMR~\cite{ohama1} and convergent-beam electron diffraction~\cite{tsuda} experiments have confirmed that NaV$_{2}$O$_{5}$ has an orthorhombic unit cell (P-phase called hereafter) with the space group of centrosymmetric $\rm{D_{2h}^{13}}$-$\rm{Pmmn}$ and lattice constants $\it{a}$ = 11.325 $\rm{\AA}$, $\it{b}$ = 3.611 $\rm{\AA}$, $\it{c}$ = 4.806 $\rm{\AA}$  under ambient conditions.
Na atoms located between V$_{2}$O$_{5}$-layers composed of two-dimensionally connected VO$_{5}$-pyramids in the $ab$-plane play a role of an electron donor~\cite{new-xray1,new-xray2}. All crystallographically equivalent V atoms are dressed a half of $\rm{3d}^{1}$ electron (V$^{4.5+}$) so that spins (S = 1/2) form a quarter-filled ladder structure along the $b$-axis. At $\it{T}_{\rm{c}}$ = 35 K, the charge-ordering~\cite{ohama1,nakao} (V$^{4.5+}$-V$^{4.5+}$$\rightarrow$V$^{4+}$-V$^{5+}$), spin-gap formation ($\Delta$ = 9.8 meV)~\cite{isobe1,fujii} and lattice dimerization characterized by the modulation wave vector ${\bf q}$=(1/2, 1/2, 1/4) ($2a \times 2b \times 4c$ in real space; C$_{1/4}$-phase called hereafter)~\cite{fujii}, take place cooperatively. 
Regarding the charge ordering pattern below $\it{T}_{\rm{c}}$, 
Seo $\&$ Fukuyama~\cite{seo-fukuyama} proposed a 2-dimensional zigzag type charge ordering model in the $ab$-plane based on the Hartree approximation by taking both on-site and inter-site Coulomb interactions into account. 
Recently, Nakao $et$ $al.$~\cite{nakao} carried out anomalous x-ray scattering experiments to selectively identify V$^{4+}$ and V$^{5+}$ ions and directly evidenced for the zigzag type charge order realized in the $ab$-plane as theoretically proposed. A full structure analysis of both atomic displacement and charge order modulation with ${\bf q}$=(1/2, 1/2, 1/4) in the low-temperature phase is very recently accomplished by x-ray scattering~\cite{nakao,sawa}. However, any microscopic mechanism to stabilize the quadrupling of its unit cell along the $c$-axis is not yet known. The pressure application will be effective for providing a key information to clarify such an origin of the long period lattice and charge modulation. This is one of purposes of the present high-pressure x-ray diffraction experiments.

Recently, Ohwada $et$ $al.$~\cite{ohwada1} have discovered a new high pressure phase which has a $2a \times 2b \times 1c$ superlattice (C$_{0}$-phase) led by the C$_{1/4}$-to-C$_{0}$ phase transition taking place at 0.92 GPa ($T$ = 8 K). On the other hand, dielectric measurements precisely performed along the $c$-axis~\cite{sekine} indicated the existence of an intermediate phase in a pressure range of 0.5 GPa $<$ $P$ $<$ 1.0 GPa at temperature $T$ = 20 K - 30 K. The second purpose of the present synchrotron x-ray diffraction experiments at low temperatures and high pressure (LT-HP) is to clarify such an intermediate phase from a microscopic structural points of view.

%%
%% Experimental
%%
For an overall survey of superlattice reflection in reciprocal space, an x-ray oscillation photographic method was first employed on Micro Powder Diffractometer (MPD)~\cite{fujiwara} equipped with an imaging plate (IP) and with an x-ray CCD system~\cite{CCD} at Photon Factory BL-1B. Wavelength was tuned to 0.6888 $\rm{\AA}$ (18 keV) with a Si(111) double-crystal monochromator. For high resolution measurement to obtain more detailed information, a detector method was employed on a four-circle diffractometer at BL-4C (PF) and BL02B1 (SPring-8). X-ray wavelength monochromatized in the same way as that of BL-1B (PF) was 0.6888 $\rm{\AA}$ at PF and 0.413 $\rm{\AA}$ (30 keV) at SPring-8. For the LT-HP experiments, a He-gas driven diamond anvil cell (DAC) was mounted on a closed-cycle He-gas refrigerator. Pressure was generated in a DAC using a 4:1 mixture of methanol:ethanol pressure transmitting media and was calibrated from a lattice constant of NaCl~\cite{nacl} enclosed  with the specimen in the DAC. The sample temperature was monitored with a Au-0.07$\%$Fe chromel thermocouple at BL-1B (PF) and a Si-diode sensor at BL-4C (PF) and BL02B1 (SPring-8) directly attached to the DAC surface. We used as-grown high quality single crystals of NaV$_{2}$O$_{5}$ with a typical size of 200 $\mu m$$\times$400 $\mu m$$\times$50 $\mu m$ ($a$$\times$$b$$\times$$c$) grown by the same method as previously reported~\cite{isobe2}.

%%
%% Result and discussion
%%
%IP result
%%%
%%% Figure 1.
%%%
\begin{figure}[t]
\begin{center}
\epsfig{file=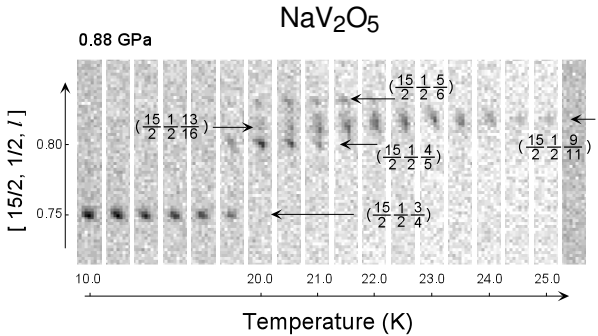,width=.45\textwidth}
\end{center}
\caption[ip]{Temperature dependence of a position of superlattice reflections (15/2, 1/2 $l$) in the region recorded on IP oscillation photographs at 0.88 GPa. Several intermediate phases with $l$ = $1/4$, $1/5$, $3/16$ and $2/11$ are systematically observed.}
\label{fig1}
\end{figure}
Figure~\ref{fig1} shows the oscillation photographs representing temperature dependence of positions of superlattice reflections along the direction of [15/2, 1/2, $l$] in selected range of reciprocal space at a fixed pressure of 0.88 GPa. All recorded reflections can be completely indexed as indicated in the figure with the aid of data processing system DENZO~\cite{DENZO}. With increasing temperature the superlattice reflection ($15/2,1/2,3/4$) disappears at 19 K while a new reflection ($15/2,1/2,4/5$) appears. Furthermore, another new reflection ($15/2,1/2,13/16$) appears at 20.0 K and gradually shifts to the position ($15/2,1/2,9/11$) between 21.5 K and 23.0 K. One can notice that all phases have the $2a \times 2b \times zc$ type superlattices whose modulation wave vector along the $c$-axis can be expressed as q$_{c}$ ( = 1/z) = $1/4$ $\rightarrow$ $1/5$ $\rightarrow$ $3/16$ $\sim$ $2/11$ in a reciprocal lattice units (r.l.u.). The superlattice with q$_{c}$ = 1/6 is also observed in a limited temperature range.
%
%Counter Results
%

%%%
%%% Figure 2.
%%%
\begin{figure}[p]
\begin{center}
\epsfig{file=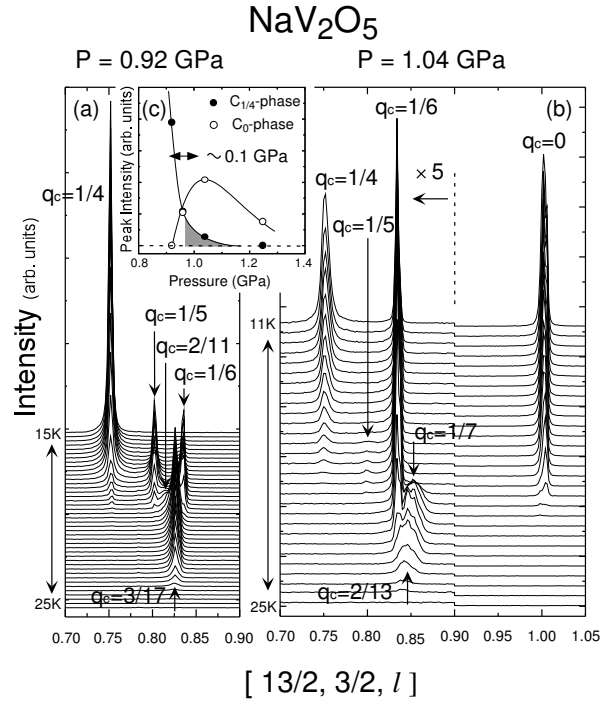,width=.45\textwidth}
\end{center}
\caption[ip]{Temperature dependence of the superlattice reflection experimentally scanned along the [13/2,3/2,$l$] direction with high resolution at (a) 0.92 GPa and (b) 1.04 GPa. (c) Peak intensities of both the C$_{1/4}$-and C$_{0}$-phases are shown to estimate the pressure distribution.}
\label{fig2}
\end{figure}
\begin{figure}[b]
\begin{center}
\psfig{file=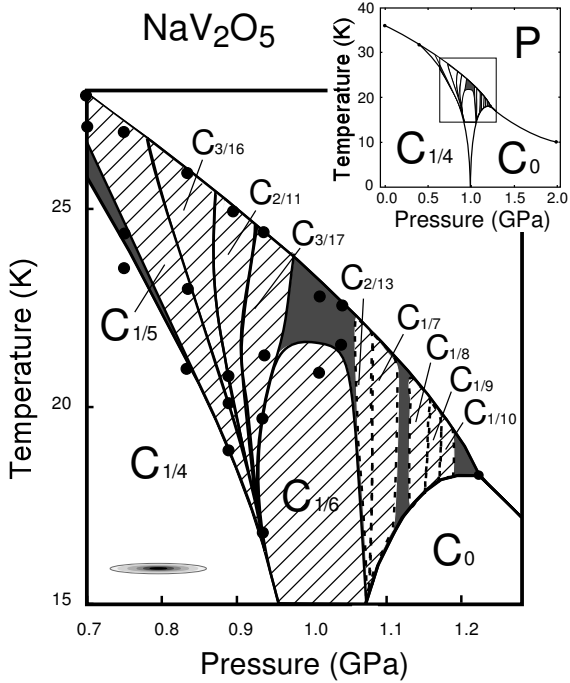,width=.45\textwidth}
\end{center}
\caption[ip]{Experimentally observed P-T phase diagram of NaV$_{2}$O$_{5}$. The hutched area shows commensurate phases unambiguously identified while the shaded area indicates more complicated higher-order commensurate or incommensurate phases unresolved by the present P-T resolution, which is represented with an ellipse at the left bottom corner. The phase boundary experimentally not clearly confirmed is drawn with dotted lines.}
\label{fig3}
\end{figure}

Figure~\ref{fig2}(a) displays temperature dependence of high-resolution diffraction profiles observed along [13/2,3/2,$l$] direction at 0.92 GPa with the counter method. In this figure a few more reflections are newly observed and one can clearly see a series of superlattice reflections with q$_{c}$ = $1/4$, $1/5$, $1/6$, $2/11$ and $3/17$ which systematically appear and disappear as a function of temperature. Figure~\ref{fig2}(b) shows the similar data observed at 1.04 GPa. At the lowest temperature 11 K, three phases with q$_{c}$ = $1/4$, $1/6$ and $0$ coexist. As temperature increases, two phases with q$_{c}$ = $1/4$ and $0$ disappear while the other phase with q$_{c}$ = $1/5$ emerges in a limited temperature range. At higher temperatures, the phases with q$_{c}$ = $2/13$ and $1/7$ show up and the three peaks shift to the more complicated higher order commensurate position. Some phases are found to coexist at a certain temperature range as seen in Fig.~\ref{fig2}(a) and (b). Such a coexistence may result from the pressure distribution in the DAC caused by the solidified pressure media. The pressure distribution can be estimated as follows: Fig.~\ref{fig2}(c) shows pressure dependence of peak intensities of the C$_{1/4}$- and C$_{0}$-phases observed at the temperature where the superlattice intensity is saturated. The two phases have a  clear phase boundary at 0.95 GPa nearly parallel to the temperature axis. From the pressure range where both phases have finite peak intensity at about 1.0 GPa, the pressure distribution is estimated as $\Delta P = 0.1$ GPa. Considering thus estimated pressure ($\Delta P = 0.1$ GPa) and temperature ($\Delta T = 0.2$ K) distributions (defined as the P-T resolution), we obtain the P-T phase diagram of NaV$_{2}$O$_{5}$ as shown in Fig.~\ref{fig3}. All phases are expressed with a characteristic modulation wave number along the $c$-axis q$_{c}$ as C$_{q_{c}}$, i.e., the C$_{1/4}$ for q$_{c}$ = 1/4 and the C$_{0}$ for q$_{c}$ = 0. 
%%%
%%% Figure 3.
%%%
We found some more phases whose boundaries could not be determined. Such unidentified phase regions are represented by dotted lines. The hutched area shows the commensurate phases unambiguously identified. The shaded area shows more complicated higher-order commensurate phases or incommensurate phases which we could not identify within the present P-T resolution. The intermediate phase is no longer observed below $P$ = 0.4 GPa and above 1.25 GPa (see the inset of Fig.~\ref{fig3}). At 2.0 GPa the P-to-C$_{0}$ phase transition temperature decreases down to 10 K.

%%%
%%% Figure 4.
%%%
The observed q$_{c}$ sequences are well understood as the "devil's staircase" type sequences theoretically obtained from the simple ANNNI model proposed by Bak and von Boehm~\cite{ANNNI}. They developed theoretical calculations with two competitive inter-layer interactions between Ising spins, i.e., the nearest neighbor interaction $J_{1}>0$ (ferro) and the next nearest neighbor interaction $J_{2}<0$ (antiferro). Within a mean field approximation, they obtained a global phase diagram and found extremely complicated phase boundaries separating various kinds of commensurate phases. It is essentially important that there are only two stable ground states as $T \rightarrow 0$, i.e., $\uparrow\uparrow\uparrow\cdot\cdot\cdot$ (ferro, all up configuration, $q=0$) state for $-J_{2}/J_{1}<1/2$ and $\uparrow\uparrow\downarrow\downarrow\uparrow\uparrow\downarrow\downarrow\cdot\cdot\cdot$ (up-up-down-down configuration, $q=1/4$) state for $-J_{2}/J_{1}>1/2$. Such a transition between the two ground states as a function of $-J_{2}/J_{1}$ was experimentally observed in NaV$_{2}$O$_{5}$ which undergoes the C$_{1/4}$-to-C$_{0}$ phase transition at 0.92 GPa~\cite{ohwada1}. The other characteristic theoretically proposed for devil's staircase at finite temperature is a systematic existence of modulated phases with the wave numbers q = $1/5$, $1/6$, $1/7$, $2/11$, $3/17$, $3/16$, $2/13$,... , which is also experimentally observed in a wide range of temperature and pressure in NaV$_{2}$O$_{5}$ as displayed in Fig.~\ref{fig3}. Thus the P-T phase diagram observed in NaV$_{2}$O$_{5}$ extremely well resembles the theoretically obtained "devil's staircase" with respect to the modulation wave number along the $c$-axis. As for the structure in the  $ab$-plane, the $2a \times 2b$ superstructure is well maintained. The present experiment directly probes the atomic displacements represented by q = (1/2, 1/2, q$_{c}$) but does not directly the charge ordering. However, the previous x-ray scattering experiments ~\cite{nakao,sawa} evidenced that in the C$_{1/4}$-phase the charge order and atomic displacements are modulated with the same wave vector q = (1/2, 1/2, 1/4).  In the present phase diagram, therefore, it is suggested that the charge order is also modulated with the same wave vector in each phase. Such a $2a \times 2b$ structure is stabilized by the inter-site Coulomb interactions~\cite{seo-fukuyama}.

\begin{figure}[t]
\begin{center}
\psfig{file=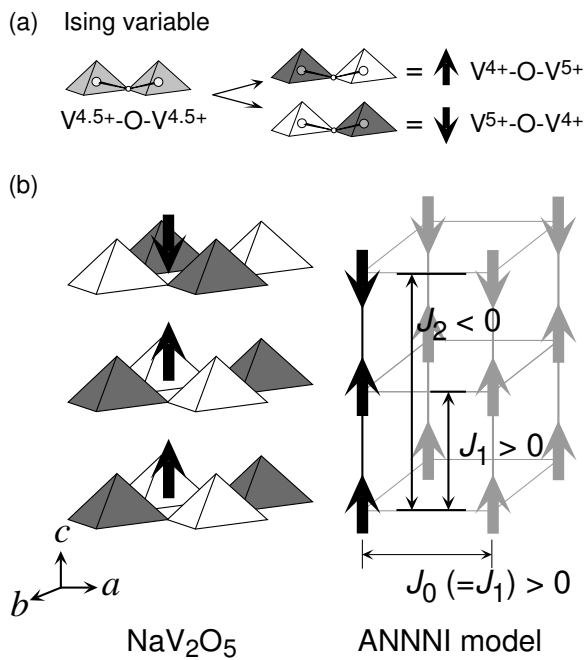,width=.45\textwidth}
\end{center}
\caption[ip]{(a) Ising spins correspond to the charge disproportionation on V-O-V rung defined in the present system. (b) Local charge arrangement of NaV$_{2}$O$_{5}$ system  along the $c$-axis and Ising spin arrangement of ANNNI system.}
\label{fig4}
\end{figure}

By analogy with the ANNNI model, the devil's staircase type behavior along the $c$-axis implies any competitive inter-layer interactions, i.e., phenomenologically $J_{1}>0$ and $J_{2}<0$ in NaV$_{2}$O$_{5}$. The inter-layer interactions must relate to the lattice constant $c$ which has compressibility as large as $\Delta$$c$/$c_{0}$ $\sim$ - 0.02 (GPa$^{-1}$)~\cite{ohwada2}.
As shown in Fig.~\ref{fig4} (a), Ising spins can be defined by the charge disproportionation on the V-O-V rung as $S$ = 1 ($\uparrow$) for the V$^{4+}$-O-V$^{5+}$ state and $S$ = -1 ($\downarrow$) for V$^{5+}$-O-V$^{4+}$ state. Local charge arrangement of NaV$_{2}$O$_{5}$ and Ising spin arrangement of ANNNI model are represented in Fig.~\ref{fig4} (b), respectively. 
It is considered that d-electrons in NaV$_{2}$O$_{5}$ are correlated in a long range with the Coulomb repulsion. $J_{2}<0$ will be understood easily by the electronic interaction; however, the $J_{1}>0$ requires more complicated mechanisms to overcome the Coulomb repulsive energy. The lattice distortion generating the polaron or the magnetic interaction between the (magnetic) spins can be one of possible origins for the attractive interaction between electrons. It is interesting to further study a microscopic mechanism of these competitive interactions and interplays between the lattice, charge and (magnetic) spin in this system.

%%
%% Conclusion
%%
%In conclusion, the "devil's staircase" type phase transition in NaV$_{2}$O$_{5}$ at LT-HP has been discovered by synchrotron radiation x-ray scattering. Successive phase transitions are found to take place and new phases with the $2a {\times} 2b {\times} zc$ type superstructures are observed. The obtained P-T phase diagram shows the feature of "devil's staircase" theoretically obtained from the ANNNI model. Observed q$_{c}$s are the wave number of atomic displacements coupled with the charge order stabilized by the same q$_{c}$. Such experimental facts imply that two competitive interactions, the nearest neighbor $J_{1}>0$ and the next nearest neighbor interactions $J_{2}<0$ along the $c$-axis, are intrinsic in NaV$_{2}$O$_{5}$. From the structural viewpoint, the contraction along the $c$-axis must drive the phase transitions associated with the charge order. A microscopic origin of the inter-layer interaction is not yet known. It is very surprising that the phase transition of such a very complicated charge-lattice-spin coupled system NaV$_{2}$O$_{5}$ may be described by the simple ANNNI model.

%%
%%Acknowledgement
%%
The instruments for the LT-HP experiments installed at PF/BL-4C were designed by Dr. Yositaka Matsusita. We thank Dr. Akihiko Fujiwara, Dr. Masakazu Nishi, Dr. Hitoshi Seo, and Prof. Hiroshi Sawa for their stimulative discussion. Experimental assistance by Mr. Yuya Katsuki and Mr. Hirondo Nakatogawa is also appreciated. One of the authors (K. O.) acknowledges support by the Japan Society for the Promotion of Science for Young Scientist. 

\vspace*{-0.5cm}
%%
%%References 
%%
%\begin{thebibliography}{99}

%\end{thebibliography}

\end{document}